\documentclass[twocolumn,aps,floatfix,showpacs,showkeys,prb,superscriptaddress]{revtex4}
\usepackage{epsfig}
\usepackage{graphicx}
\usepackage{amsmath}
\usepackage{amssymb}
\usepackage{amsfonts}
\usepackage{tabularx}
\usepackage{color}
\usepackage{dcolumn}
\usepackage{bm}

\newcommand{\ls}{$m_L/m_S$}

\begin{document}

\title{Influence of plural scattering on the quantitative determination of spin and orbital moments in electron magnetic circular dichroism measurements}

\author{J\'{a}n Rusz} 
\affiliation{Department of Physics and Materials Science, Uppsala University, Box 530, S-751 21 Uppsala, Sweden}
\affiliation{Institute of Physics, Academy of Sciences of the Czech Republic, Na Slovance 2, CZ-182 21 Prague, Czech Republic}

\author{Hans Lidbaum}
\affiliation{Department of Engineering Sciences, Uppsala University, Box 534, S-751 21 Uppsala, Sweden}

\author{Stefano Rubino}
\affiliation{Department of Engineering Sciences, Uppsala University, Box 534, S-751 21 Uppsala, Sweden}

\author{Bj\"{o}rgvin Hj\"{o}rvarsson}
\affiliation{Department of Physics and Materials Science, Uppsala University, Box 530, S-751 21 Uppsala, Sweden}

\author{Peter M. Oppeneer}
\affiliation{Department of Physics and Materials Science, Uppsala University, Box 530, S-751 21 Uppsala, Sweden}

\author{Olle Eriksson}
\affiliation{Department of Physics and Materials Science, Uppsala University, Box 530, S-751 21 Uppsala, Sweden}

\author{Klaus Leifer}
\affiliation{Department of Engineering Sciences, Uppsala University, Box 534, S-751 21 Uppsala, Sweden}

\date{\today}

\begin{abstract}
Recent quantitative measurements of the orbital to spin magnetic moment ratio \ls{} in electron magnetic circular dichroism (EMCD) experiments have given a \ls{} ratio that is larger than commonly accepted values. We demonstrate here that plural scattering may noticeably influence the \ls{} ratio. An equation is derived which describes its influence as a function of the spectral integrals of the plasmon scattering region and zero-loss peak. The influence of the electron-plasmon scattering can be removed when electron energy-loss spectra of the ionization edge are deconvoluted by the low-loss signal. For a bcc-Fe sample we obtain \ls{}$=0.04$ after plasmon removal. We conclude that the plural scattering should be considered when extracting quantitative information from EMCD measurements.
\end{abstract}

\pacs{}
\keywords{circular dichroism, transmission electron microscopy, density functional theory, dynamical diffraction theory, plural scattering, multiple inelastic scattering}

\maketitle

Electron magnetic circular dichroism (EMCD) is a young experimental technique,\cite{nature} which in principle allows to extract atom-specific spin and orbital magnetic moments from electron energy-loss spectra by applying EMCD sum rules.\cite{oursr,lionelsr} This technique, as compared to its counterpart x-ray magnetic circular dichroism (XMCD) \cite{stohr}, brings a promise of nanometer or even sub-nanometer spatial resolution \cite{emcd2nm,lsfollow} with bulk sensitivity using a standard transmission electron microscope (TEM).

Recently a few works reported the first EMCD evaluations of the orbital to spin moment ratios \ls{} --- for bcc-Fe $0.09 \pm 0.03$,\cite{lionelsr} $0.08 \pm 0.01$,\cite{recipmaps} and for hcp-Co $0.14 \pm 0.03$ \cite{schatt}). It should be noted that all these values overestimate the \ls{} ratios as measured by Chen \textit{et al.}\ \cite{chen} using XMCD (0.043 for bcc-Fe and 0.095 for hcp-Co). In this paper we demonstrate that a possible explanation of this overestimation is based on \emph{plural} scattering, \textit{i.e.}, low-loss excitations accompanying the core-level ionization processes. 


In our recent work \cite{recipmaps} we reported measurements of energy-resolved diffraction patterns on a 20 nm bcc-Fe thin film, which were acquired in two-beam (2BC) and three-beam (3BC) case orientations of the sample. Based on symmetry arguments \cite{theory,lsfollow} the 3BC is preferred and a new \emph{double difference} procedure has been introduced for extraction of the magnetic component of the signal. The electron energy-loss spectra (EELS) were in Ref.~\onlinecite{recipmaps} processed for every pixel in reciprocal space and thus evaluated by extracting data from maps of the signal. Instead of working with \ls{} values from individual pixels it is possible to use averaged spectra from the same set of data\cite{snr}, which should lead to the same results under symmetrical conditions \cite{theory}. Here, spectra were obtained from the data cube of Ref.~\onlinecite{recipmaps} (after application of point blemish removal, cross-correlation and rotation, see Refs.~\onlinecite{recipmaps,lsfollow} for details) by selecting two virtual apertures of 70$\times$70 pixels (corresponding to a square of size $G_{(200)}/2$), according to the optimum position and size determined in Ref.~\onlinecite{recipmaps}. Each spectrum was thereafter background subtracted, normalized in the post-edge region and fitted using the same models as described in Ref.~\onlinecite{lsfollow}. The method of averaged spectra improves the signal to noise ratio and is therefore preferable here for studying the influence of plural scattering.

\begin{figure}
  \includegraphics[width=8.5cm]{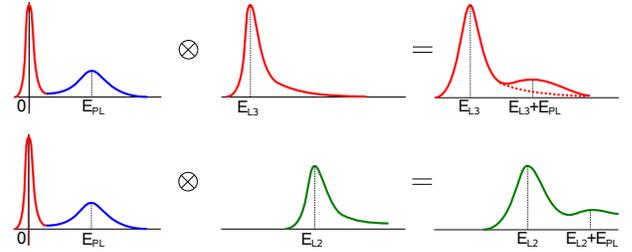}
  \caption{Illustration of the influence of plural scattering. Top: convolution of the low-loss region, consisting of zero-loss peak and plasmon peak (at $E_\text{PL}$), with a peak at $E_{L_3}$ leads to a double-peak structure. Bottom: the same procedure applied on $L_2$ peak. If $E_{L_2}$ is close to $E_{L_3}+E_\text{PL}$, then the $L_2$ peak is enhanced by the plural scattering accompanying the $L_3$ excitation.\label{fig:convolution}}
\end{figure}

It is assumed that the effect of the plural scattering can be modelled by a convolution of a net (single scattering) near-edge spectrum with the spectrum from the low-loss region,\cite{egerton} as illustrated in Fig.~\ref{fig:convolution}. An $L_3$ peak at energy-loss $E_{L_3}$ is convoluted with the low-loss spectrum, which has a double-peak structure: an intense zero-loss peak (ZLP) and a relatively weak plasmon peak at $E_\text{PL}$. Result of this convolution is again a double-peak structure with main peak at $E_{L_3}$ and an additional peak at $E_\text{PL}+E_{L_3}$. The latter appears due to double excitation events, \textit{i.e.}, when one fast electron excites a plasmon and a $L_3$ core-level transition. The energy-integral of this additional peak is proportional to the intensity of the $L_3$ peak and to the ratio $k$ of energy-integrals of plasmon peak ($A_{PL}$) and ZLP ($A_{ZLP}$), $k=A_{PL}/A_{ZLP}$. It is important to observe that if this peak overlaps with the $L_2$ edge, then the $L_2$ edge is enhanced by low-loss inelastic scattering events accompanying the $L_3$ excitation.

\begin{figure}
  \includegraphics[angle=270,width=8.5cm]{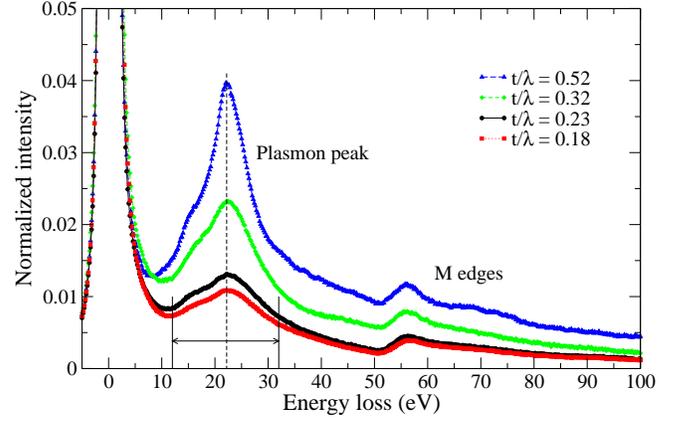}
  \caption{Low-loss EELS spectra of bcc-Fe measured at sample thicknesses $t/\lambda=0.18, 0.23, 0.32$ and $0.52$. At $t/\lambda=0.32$ the beam passes also through a V/Fe superlattice and at $t/\lambda=0.52$ additionally through MgO substrate. The zero-loss peak height was normalized to one and centered at 0 eV for all thicknesses.\label{fig:plasmon}}
\end{figure}

The low-loss spectra, measured from the bcc-Fe sample at four different thicknesses, are shown in Fig.~\ref{fig:plasmon}. Note that the peak-to-peak ratio between the plasmon peak and the ZLP is only around 1\%. However, the spectral integral of the plasmon region (within the energy-loss range 10 to 35 eV) is approximately 10\% and 12\% of the ZLP area at thicknesses $t/\lambda=0.18$ and $0.23$, respectively ($t$ is thickness and $\lambda$ is the mean free path\footnote{The mean free path $\lambda$ for 300 keV electrons in Fe is roughly around 100 nm.\cite{egerton} The pure ZLP was obtained by using different peak fitting models available in the Digital Micrograph\texttrademark software, such as reflected tail or combination of Gaussian/Lorentzian (resulting in a FWHM of approximately 2 eV).}). Note that the $E_{PL}$ is approximately at 22 eV with full-width at half-maximum above 20 eV, Fig.~\ref{fig:plasmon}, and the energy difference of $L_3$ and $L_2$ edges in Fe is cca 13 eV. Therefore in bcc-Fe the plasmon excitations accompanying the $L_3$ excitation indeed overlap with the $L_2$ peak.

A simplified estimation of the influence of plural scattering on the \ls{} ratio can be obtained in the following way: the convolution of near-edge spectra with low-loss spectra enhances the $L_2$ peak proportionally to the ratio $k$ and the intensity of the $L_3$ edge. \footnote{More precisely one should take the energy-integral of the plasmon peak within the suitable energy range, over which the near-edge spectra are peak-fitted, here energy-loss of ca.\ 10--35 eV.} Assuming that symmetry conditions for detector positions required by the EMCD sum-rules \cite{oursr} are fulfilled, the spectra at detector positions + and -- are schematically given by:
\begin{eqnarray}
  \frac{\partial^2 \sigma(+)}{\partial \Omega \partial E} & = & \overbrace{N_1+M_1}^{L_3} + [\overbrace{N_2+M_2}^{L_2\text{ net}}+\overbrace{k(N_1+M_1)}^{\text{plural scattering}}] \\
  \frac{\partial^2 \sigma(-)}{\partial \Omega \partial E} & = & \underbrace{N_1-M_1}_{L_3} + \underbrace{[N_2-M_2+k(N_1-M_1)]}_{L_2\text{ total}}
\end{eqnarray}
where $N,M$ denote the non-magnetic and magnetic component of the signal, indices $1$ and $2$ correspond to lower and higher energy-loss edges, respectively. For late transition metals $M_1$ and $M_2$ have typically opposite signs. The EMCD difference spectrum is, consequently:
\begin{equation} \label{eq:dif}
  \frac{\partial^2 \Delta \sigma}{\partial \Omega \partial E} = 2M_1 + [2M_2+2kM_1].
\end{equation}
The effect of plural scattering does not cancel out when taking the difference spectrum because of the different intensity of the magnetic terms for the two detector positions.

When we insert the energy integrated values for individual edges into the sum-rule expression for the \ls{} ratio,\cite{oursr} we obtain after some algebra the relation between `clean' \ls{} value, $u$, and the `convoluted' one, $\tilde{u}$:
\begin{eqnarray}
  \tilde{u} & = & \frac{(2L-1)u + L(L-1)k/3 + kLu}{2L-1 - kL - 3kuL/(L-1)} \label{eq:exact1}\\
         u  & = & \frac{(2L-1)\tilde{u} - L(L-1)k/3 - kL\tilde{u}}{2L-1 + kL + 3k\tilde{u}L/(L-1)} \label{eq:exact2}
\end{eqnarray}
where $L$ orbital quantum number of final states ($L=2$ for $p \to d$ and $L=3$ for $d \to f$). At $u=\frac{1-L}{3}$ the plural scattering has no influence on \ls{} ratio, \textit{i.e.}, $u=\tilde{u}$. It can be shown that the value $u=\frac{1-L}{3}$ corresponds to $M_1=0$. This explains the $k$-independence of $\tilde{u}$, see Eq.~(\ref{eq:dif}).

By expanding into first order in $k$ and omitting a term proportional to the square of the clean \ls{} ratio (negligible for late transition metals) we obtain simpler equations for $L_{2,3}$ edges:
\begin{equation} \label{eq:enh}
  \left. \frac{m_L}{m_S} \right|_{\text{conv.}} \approx \left. \frac{m_L}{m_S} \right|_{\text{clean}} \left( 1 + \frac{4k}{3} \right) + \frac{2k}{9}
\end{equation}
or correspondingly
\begin{equation} \label{eq:enhINV}
  \left. \frac{m_L}{m_S} \right|_{\text{clean}} \approx \left. \frac{m_L}{m_S} \right|_{\text{conv.}} \left(1 - \frac{4k}{3}\right) - \frac{2k}{9}.
\end{equation}

The second term in Eq.~(\ref{eq:enh}) predicts that if $k=10\%$, then the \ls{} ratio obtained by straightforward application of sum rules is enhanced by 0.022. The factor in parentheses has a smaller effect under the assumptions made here, giving an enhancement by ca.\ 13\%. Applying Eq.~(\ref{eq:enh}) to the value obtained in Ref.~\onlinecite{recipmaps} $(m_L/m_S)_\text{conv.}=0.08$ leads to $(m_L/m_S)_\text{clean}=0.05$, which is a considerable change.

\begin{figure}
  \includegraphics[angle=0,width=7.5cm]{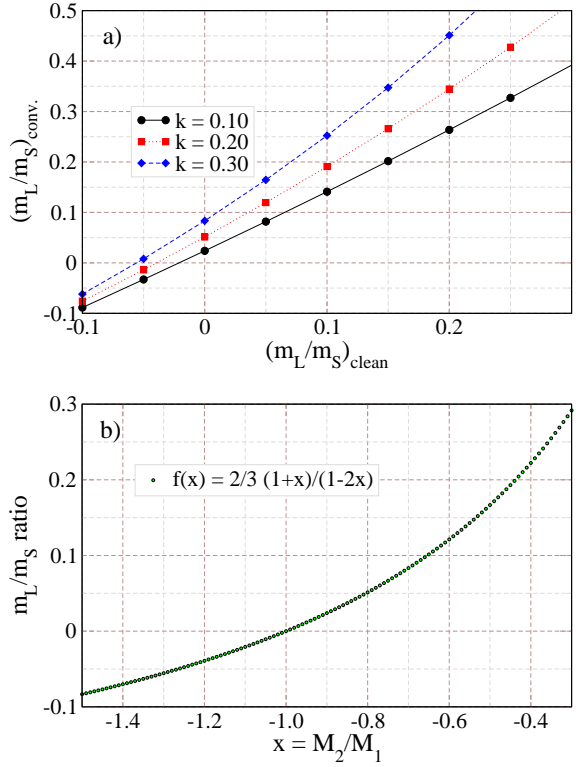}
  \caption{(color online) a) Qualitative effect of plural scattering for different values of the area ratio $k$ between the plasmon region and ZLP, Eqs.~(\ref{eq:exact1}) and (\ref{eq:exact2}). b) \ls{} value as a function of $x=M_2/M_1$ (ratio of the integrated dichroic signals from $L_2$ and $L_3$ edges).\label{fig:enh}}
\end{figure}

Figure~\ref{fig:enh} shows the prediction of Eq.~(\ref{eq:enh}) for different strengths of the plasmon region, $k$, thus revealing the relation between convoluted and deconvoluted (clean) \ls{} ratio. 
From the raw spectra, convoluted values of the areas are available, namely $\tilde{M}_1$ and $\tilde{M}_2=M_2+kM_1$. When also $k$ is known, from Fig.~\ref{eq:enh} `clean' \ls{} ratios can be estimated. According to the EMCD sum rules,\cite{oursr,lionelsr} the clean or convoluted \ls{} ratio is a function of $x=M_2/M_1$ or $\tilde{M}_2/\tilde{M}_1$, respectively.
For the $L_{2,3}$ edges this gives $m_L/m_S=\frac{2}{3} (1+x)/(1-2x)$, which is shown in Fig.~\ref{fig:enh}b.

To study the effect on a quantitative level, we performed a plasmon removal procedure on our experimental data taken in the 3BC orientation. Fourier-ratio deconvolution\cite{egerton,wang,jodeconv} was performed in Digital Micrograph\texttrademark, using a zero-loss modifier as reconvolution method. To test the sensitivity of the extraction of the data we positioned the set of boxes first at the Thales circle positions,\cite{nature} and subsequently at the optimal positions with maximum relative dichroic signal.\cite{recipmaps} Each set comprises of four spectra, one in each quadrant of the diffraction plane. It should be noted that the ZLP could not be acquired under exactly the same illumination conditions as the near-edge spectra due to very high intensity of the electron beam in the telefocus mode\cite{lsfollow}, which could damage the CCD camera.

\begin{figure}
  \includegraphics[width=7.5cm]{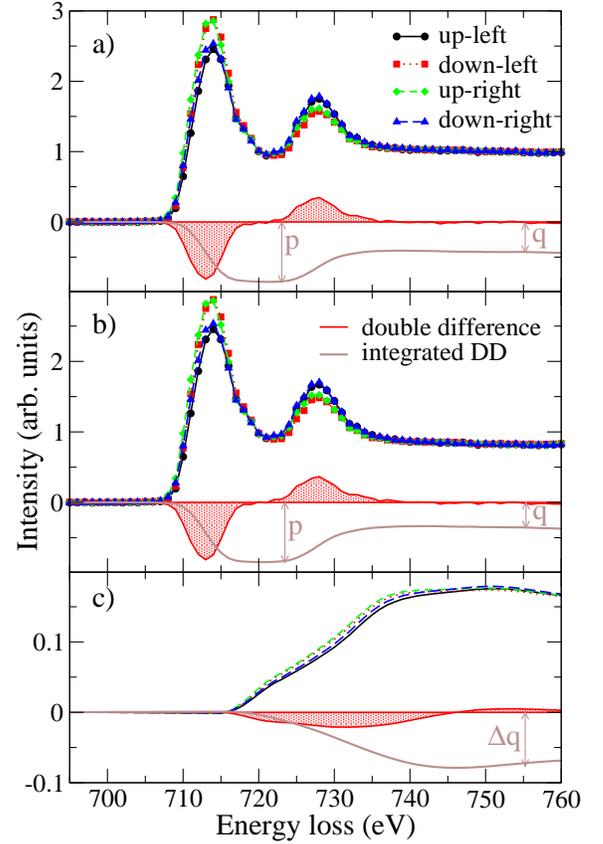}
  \caption{(color online) Influence of the plasmon removal on energy-loss and dichroic spectrum, see text for details. a) Averaged normalized spectra from all four quadrants (see legend); b) the same spectra after plasmon removal; c) The difference of spectra from panels a) and b). In all panels: the red curve with hatching relates to the double difference signal (DD) and the grey curve to its energy-integral. Values $p$ and $q$ enter the sum rules expression of Ref.~\onlinecite{chen}.\label{fig:spec}}
\end{figure}

In Fig.~\ref{fig:spec}a we show the averaged spectra from all four quadrants at the optimal positions. Optimum positioning of the box in combination with the double difference procedure results in a strong EMCD signal, reaching almost $30\%$ of the $L_3$ edge peak height. Analyzing this signal by means of peak-fitting as described in Ref.~\onlinecite{lsfollow} we obtain $(m_L/m_S)_\text{conv.}=0.09$, in agreement with the value of Ref.~\onlinecite{recipmaps} where the value was extracted from \ls{} ratio maps. Figure~\ref{fig:spec}b shows the same spectra after plasmon removal using the low energy-loss spectrum measured at $t/\lambda=0.23$, which corresponds to the sample thickness at which the data were obtained. The resulting change is relatively weak, as could be expected taking into account the small plasmon area as compared to the ZLP. However, the higher energy-loss tails end at an intensity of approximately 0.85 instead of unity as in the normalized spectra of Fig.~\ref{fig:spec}a.

Figure~\ref{fig:spec}c displays the effect of plasmon removal as a difference between spectra from panel a) and panel b). The onset of the enhancement starts approximately at 10 eV above the onset of the $L_3$ edge, in agreement with the low energy-loss spectra, Fig.~\ref{fig:plasmon}, where the 0--10 eV region is dominated by the ZLP. At about 723 eV there is a shoulder which originates from the same feature of the plasmon peak at around 15 eV. The broad humps around 735 eV and 750 eV are derived from $L_3$ and $L_2$ edges smeared by the wide plasmon peak. These features are, compared to $L_{2,3}$ peak positions, shifted up in energy by approximately 20--25 eV. This shift is associated to the position of the plasmon peak, around 22 eV above zero energy-loss.

\begin{table}[t]
  \begin{tabular*}{\columnwidth}{@{\extracolsep{2mm}}l*{4}{r}}
\hline\hline
             & \multicolumn{2}{c}{Thales circle} & \multicolumn{2}{c}{Optimum pos.} \\
 Orientation & raw data      & deconv.        & raw data        & deconv.          \\
\hline 

    Up-down, right & 0.07 & 0.02 & 0.07 & 0.02 \\
    Up-down, left  & 0.10 & 0.06 & 0.10 & 0.05 \\
    Double difference   & 0.09 & 0.04 & 0.09 & 0.04 \\

\hline \hline
  \end{tabular*}
  \caption{Summary of the \ls{} values obtained in 3BC geometry from raw spectra and from spectra after plasmon removal, both at the optimal placement and Thales circle positions, see text for details.\label{tab:ls}}
\end{table}

Since the plasmon removal will alter also the double-step background,\cite{lsfollow} which in turn can influence the $L_3$ edge area, we applied the peak fitting procedure described in Refs.~\onlinecite{recipmaps,lsfollow} to deconvolved spectra, Fig.~\ref{fig:spec}b. The results are summarized in the Table~\ref{tab:ls}. There we show \ls{} values in 3BC orientation, using a square integration box of $0.5G_{200} \times 0.5G_{200}$ at optimal placement and Thales circle positions, prior to and after the plasmon removal. The up-down difference for left and right side refers to the use of a horizontal mirror axis.\cite{lsfollow} The double difference combines both horizontal and vertical mirror axes.\cite{recipmaps} It effectively cancels out deviations in the \ls{} ratio, induced by asymmetry and misalignments in the experiment.\cite{lsfollow,theory} After plasmon removal we obtained for bcc-Fe a \ls{} ratio of $0.04 \pm 0.01$ in this setup.

We checked the method on the only other EMCD \ls{} value available in the literature, the hcp-Co\cite{schatt}. We estimate that for a 18~nm thick hcp-Co sample, the value for $k$ (in the same energy range 10--35~eV) would be $\approx0.1$.
Applying the Eq.~\ref{eq:exact2} on value (\ls{})$_\mathrm{conv}=0.14$ gives (\ls{})$_\mathrm{clean}=0.10$, which agrees with XMCD value 0.095\cite{chen} within its error bar.

We have demonstrated that the plural scattering has a non-negligible influence on quantitative information extracted from EMCD spectra. Therefore, a measurement of the low energy-loss spectra and subsequent plasmon removal from the near-edge spectra is a necessary part of the quantitative EMCD analysis. A qualitative formula for estimation of the error introduced by plural scattering was derived. The correction is found to scale approximately linearly with the ratio of spectral integrals of the plasmon region and ZLP. By applying the plasmon removal procedure to spectra measured in Ref.~\onlinecite{recipmaps} we obtain $m_L/m_S=0.04 \pm 0.01$, which is in good agreement with values 0.043 and 0.044 obtained by XMCD \cite{chen} and gyromagnetic ratio\cite{reck}, or 0.062 by neutron scattering experiments\cite{landolt} respectively.

This work was supported through the Swedish Research Council (VR), Knut and Alice Wallenberg foundation (KAW), G\"{o}ran Gustafsson foundation, and STINT.


\begin{thebibliography}{99}

\bibitem{nature} P. Schattschneider, S. Rubino, C. H\'{e}bert, J. Rusz, J. Kune\v{s}, P. Nov\'{a}k, E. Carlino, M. Fabrizioli, G. Panaccione, and G. Rossi, Nature {\bf 441}, 486 (2006).

\bibitem{oursr} J. Rusz, O. Eriksson, and P. Nov\'{a}k, P. M. Oppeneer, Phys. Rev. B {\bf 76}, 060408(R) (2007).

\bibitem{lionelsr} L. Calmels, F. Houdellier, B. Warot-Fonrose, C. Gatel, M. J. H\"{y}tch, V. Serin, E. Snoeck, and P. Schattschneider, Phys. Rev. B {\bf 76}, 060409(R) (2007).

\bibitem{stohr} J. St\"{o}hr, Y. Wu, B. D. Hermsmeier, M. G. Samant, G. R. Harp, S. Koranda, D. Dunham, and B. P. Tonner, Science {\bf 259}, 658 (1993).

\bibitem{emcd2nm} P. Schattschneider, M. St\"{o}ger-Pollach, S. Rubino, M. Sperl, Ch. Hurm, J. Zweck, and J. Rusz, Phys. Rev. B {\bf 78}, 104413 (2008).

\bibitem{lsfollow} H. Lidbaum, J. Rusz, S. Rubino, A. Liebig, B. Hj\"{o}rvarsson, P. M. Oppeneer, E. Coronel, O. Eriksson, and K. Leifer, submitted, arXiv:0908.3963v1.

\bibitem{recipmaps} H. Lidbaum, J. Rusz, A. Liebig, B. Hj\"{o}rvarsson, P. M. Oppeneer, E. Coronel, O. Eriksson, and K. Leifer, Phys. Rev. Lett. \textbf{102}, 037201 (2009).

\bibitem{schatt} P. Schattschneider, S. Rubino, M. St\"{o}ger-Pollach, C. H\'{e}bert, J. Rusz, and L. Calmels, J. Appl. Phys. \textbf{103}, 07D931 (2008).

\bibitem{chen} C. T. Chen, Y. U. Idzerda, H-J. Lin, N. V. Smith, G. Meigs, E. Chaban, G. H. Ho, E. Pellegrin, and F. Sette, Phys. Rev. Lett. \textbf{75}, 152 (1995).

\bibitem{theory} J. Rusz, submitted to Phys. Rev. B.

\bibitem{snr} J. Verbeeck, C. H\'{e}bert, P. Schattschneider, S. Rubino, P. Nov\'{a}k, J. Rusz, F. Houdellier, and C. Gatel, Ultramicroscopy \textbf{108}, 865 (2008).

\bibitem{egerton} R. F. Egerton, \textit{Electron Energy-Loss Spectroscopy in the Electron Microscope}, (Plenum Press, New York, ed. 2, 1996).

\bibitem{wang} F. Wang, R. Egerton, M. Malac, Ultramicroscopy \textbf{109}, 1245 (2009).

\bibitem{jodeconv} J. Verbeeck, and G. Bertoni, Ultramicroscopy, in press.

\bibitem{reck} R. A. Reck, and D. L. Fry, Phys. Rev. {\bf 184}, 492 (1969).

\bibitem{landolt} M. B. Stearns, in \textit{Landolt-B\"{o}rnstein Numerical Data and Functional Relationships in Science and Technology}, (Springer, Berlin, 1986), Group 3, Vol. 19, Pt. a , p.53.

\end{thebibliography}
\end{document}